\documentclass[conference]{IEEEtran}

\usepackage{cite}
\usepackage{amsmath,amssymb,amsfonts}
\usepackage{algorithmic}
\usepackage{graphicx}
\usepackage{textcomp}
\usepackage{xcolor}
\usepackage{booktabs}
\usepackage{tcolorbox}
\usepackage{xspace}
\usepackage{tikz}
\usepackage{url}
\usepackage{balance}

\newtcolorbox{resultbox}[1][]{
    colback=gray!10,
    colframe=gray!60,
    boxrule=0pt,
    leftrule=2.5pt,
    arc=0mm,
    left=5pt, right=4pt,
    top=3pt, bottom=3pt,
    before skip=6pt,
    after skip=6pt,
    #1
}

\usetikzlibrary{fit,positioning,shapes.geometric,arrows.meta,calc}
\input{shapes/database}
\tikzstyle{double_arrow} = [thick,<->,>=stealth]
\tikzstyle{arrow} = [thick,->,>=stealth]
\tikzstyle{dash_arrow} = [thick, dashed, ->, >=stealth]
\tikzstyle{dash_double_arrow} = [thick, dashed, <->, >=stealth]
\tikzstyle{start} = [circle, draw, fill=black, minimum size=0.6cm, align=center, font=\small\bfseries, text=white]
\tikzstyle{end}   = [circle, draw, fill=black, minimum size=0.6cm, align=center, font=\small\bfseries, text=white, double=white, double distance=2pt, thick]
\tikzstyle{box} = [draw, rectangle, rounded corners, minimum height=0.8cm, align=center, font=\small, inner sep=0.2cm]
\tikzstyle{decision} = [draw, diamond, aspect=1, minimum size=0.7cm, inner sep=0pt]
\tikzstyle{subbox} = [draw, rectangle, dotted, thick, rounded corners, inner sep=0.4cm]
\tikzstyle{bar} = [draw, rectangle, fill=black, minimum width=1.5cm, minimum height=0.1cm, inner sep=0pt]
\tikzstyle{bar_vertical} = [draw, rectangle, fill=black, minimum width=0.1cm, minimum height=1.8cm, inner sep=0pt]
\tikzstyle{time_event} = [minimum size=0.8cm, inner sep=0pt, path picture={\draw[thick] (path picture bounding box.north west) -- (path picture bounding box.north east) -- (path picture bounding box.south west) -- (path picture bounding box.south east) -- cycle;}]
\tikzstyle{flow_final_node} = [circle, draw, thick, minimum size=0.6cm, inner sep=0pt, path picture={\draw[thick] (path picture bounding box.north west) -- (path picture bounding box.south east); \draw[thick] (path picture bounding box.south west) -- (path picture bounding box.north east);}]
\tikzstyle{pod} = [draw, rectangle, rounded corners, minimum height=0.8cm, align=center, font=\small, inner sep=0.2cm]

\tikzstyle{lifeline} = [draw, rectangle, fill=blue!5, minimum height=0.6cm, minimum width=2.5cm, align=center, rounded corners=2pt]
\tikzstyle{activation} = [draw, rectangle, fill=white, minimum width=0.2cm, inner sep=0pt]
\tikzstyle{seq_message} = [->, >=stealth, thick]
\tikzstyle{seq_reply} = [->, >=stealth, dashed, thick]
\tikzstyle{alt_box} = [draw, rectangle, dashed, thick, rounded corners=2pt]

\newcommand{\restberus}[0]{RESTberus\xspace}
\newcommand{\restgym}[0]{RESTgym\xspace}

\newcommand{\deeprest}[0]{DeepREST\xspace}
\newcommand{\morest}[0]{MoREST\xspace}
\newcommand{\restest}[0]{RESTest\xspace}
\newcommand{\restler}[0]{RESTler\xspace}
\newcommand{\resttestgen}[0]{RestTestGen\xspace}
\newcommand{\schemathesis}[0]{Schemathesis\xspace}

\begin{document}

\title{RESTberus: Trace-Aware Benchmarking of Automated REST API Testing Tools Across Microservice Dependencies}

\author{
   \IEEEauthorblockN{John Toniutti}
   \IEEEauthorblockA{\textit{University of Udine} -- Italy \\ toniutti.john@spes.uniud.it}
   \and
   \IEEEauthorblockN{Massimiliano Baldo}
   \IEEEauthorblockA{\textit{IMT Lucca} -- Italy \\ massimiliano.baldo@imtlucca.it}
   \and
   \IEEEauthorblockN{Marino Miculan}
   \IEEEauthorblockA{\textit{University of Udine} -- Italy \\ marino.miculan@uniud.it}
   \and
   \IEEEauthorblockN{Vincenzo Riccio}
   \IEEEauthorblockA{\textit{University of Udine} -- Italy \\ vincenzo.riccio@uniud.it}
}

\maketitle

\begin{abstract}
Existing REST API testing benchmarks focus on isolated services, overlooking failures arising from interactions with downstream dependencies.
In microservice systems, response bodies and entry-point operations alone cannot reliably distinguish failures, compromising tool comparisons.
We present \restberus, a Kubernetes-based benchmarking platform for black-box REST API testing tools on deployed microservice systems.
By complementing API-level evidence with distributed traces, \restberus can distinguish failures manifesting in the target service from those manifesting in downstream dependencies.
Across six tools on two microservice benchmarks, our trace-aware deduplication strategy achieves 87.1\% accuracy, outperforming baselines from prior work by 10--13 percentage points.
Our empirical study shows that all tools expose downstream failures, accounting for 11.7\% of estimated unique failures on average, which isolated-service benchmarks cannot expose.
On single-service benchmarks, \restberus produces statistically equivalent results to \restgym, demonstrating that integration-level assessments can preserve compatibility with existing tools and comparability with prior evaluations.

\end{abstract}

\begin{IEEEkeywords}
REST APIs, Microservice Testing, Benchmarking Infrastructure, Distributed Tracing, Integration Testing
\end{IEEEkeywords}

\section{Introduction}

REST APIs are widely used to build the backends of modern web and cloud applications, especially in microservice architectures, where they allow clients to access backend functionality and services to communicate with each other~\cite{GolmohammadiAmid2024TRAA}.
Ensuring their reliability has motivated the development of several automated black-box testing tools, which generate requests from OpenAPI specifications to exercise API operations and expose failures.
As these tools evolve, e.g., including deep learning approaches~\cite{deeprest}, reproducible benchmarking is essential to compare their effectiveness under controlled conditions.

However, existing benchmarking infrastructures focus on isolated services.
In particular, \restgym~\cite{CorradiniDavide2025RAFI} provides a reproducible environment for comparing REST API testing tools, but evaluates individual services by observing HTTP responses at the API boundary and estimating unique failures through response-body similarity.
This approach overlooks the challenges introduced by distributed microservice architectures.

In fact, a single API request within a microservice system may trigger calls to several downstream services before a response is returned.
An HTTP 5xx response observed at the API boundary may therefore reflect a failure manifested either in the target service or in a downstream dependency.
Moreover, intermediate services may return the same generic error response for different failures, while the same downstream failure may be exposed through different entry-point operations.
Consequently, API-boundary observations alone cannot distinguish target-side from downstream failures, and grouping failures solely by response similarity or API operation may produce misleading estimates of unique failures.

Moving from isolated services to microservices requires more sophisticated orchestration and monitoring, but this technological leap should not come at the cost of comparability with established benchmarks.

MacroHive~\cite{GiamatteiLuca2024AFaR} successfully extends automated REST API testing to deployed microservice systems through dedicated test generation and monitoring of inter-service interactions.
Although its approach is compared to testing tools, its empirical comparison does not apply the same internal observability and failure-analysis methodology to all compared tools.

This leaves an important gap: \emph{how can we benchmark existing REST API testing tools on real microservice deployments, measure failures across service boundaries, and preserve comparability with established benchmarks?}

We address this gap with \restberus, a Kubernetes-based benchmarking platform that evaluates existing black-box REST API testing tools without modifying their test-generation logic.
\restberus automatically deploys each target service with its downstream dependencies, isolates experimental runs, controls random seeds and interaction budgets, and collects API-boundary interactions, code coverage, and distributed traces.
It introduces a trace-aware deduplication strategy to distinguish failures manifesting in the target service from those manifesting downstream.
These capabilities are implemented within the benchmarking infrastructure, allowing existing approaches to benefit from richer evaluation without being redesigned.

We evaluate \restberus using six testing tools on Train Ticket and PetClinic and nine single-service benchmarks from \restgym.
Our proposed deduplication strategy achieves 87.1\% deduplication accuracy, outperforming existing strategies by up to 13.0 percentage points.
It also achieves the highest observed accuracy (92.2\%) on human-annotated failure pairs.
All six testing tools expose downstream failures, which account for 11.7\% of their unique failures on average, demonstrating the importance of accounting for downstream failures.
Finally, equivalence testing shows that \restberus preserves the evaluation results of \restgym on single-service benchmarks.
Together, these findings demonstrate that benchmarking can move beyond API-boundary observations without sacrificing compatibility with existing tools or comparability with prior results.

The main contributions of this paper are:
\begin{itemize}
\item \textbf{Integration-level microservice benchmarking.}
We introduce \restberus to evaluate black-box REST API testing tools on deployed service chains, with controlled execution and integrated telemetry collection.
\item \textbf{Trace-aware failure analysis.}
We introduce attribution of target and downstream failure manifestations and a deduplication strategy that combines distributed traces with API-level evidence, improving upon existing approaches.
\item \textbf{Empirical evidence and backward compatibility.}
We evaluate six testing tools on distributed benchmarks, quantify the contribution of downstream failures, assess deduplication against LLM and human reference labels, and establish practical equivalence with \restgym on single-service applications.
\end{itemize}

\section{Background} \label{sec:background}

\subsection{REST APIs and Automated Black-Box Testing}

REST is an architectural style for web services, commonly implemented over HTTP, in which clients manipulate server resources by issuing requests with a method, a resource path, and an optional payload~\cite{fielding2000architectural}.
Servers report outcomes with standardized status codes, where the 5xx class indicates server-side errors~\cite{rfc9110}.

REST APIs are commonly documented with machine-readable specifications such as the OpenAPI Specification~\cite{openapi_spec}, which define operations, their inputs, and their expected responses.
This contract enables black-box testing tools to exercise a service through its interface, without access to its source code~\cite{swebok_v4}, by deriving valid requests and sequences of HTTP interactions.
Without access to internal program state, these tools rely on HTTP status codes as test oracles and commonly treat unexpected 5xx responses as failures.

\begin{figure}[t!]
  \centering
  \resizebox{0.85\columnwidth}{!}{
    \begin{tikzpicture}[
    comp/.style={draw, rectangle, rounded corners, minimum width=2.4cm, minimum height=0.8cm, align=center, font=\small},
    scope/.style={draw, rounded corners, inner sep=#1},
  ]

  \node[comp, fill=blue!5]   (api)   at (0,2.0) {API layer};
  \node[comp, fill=blue!5]   (logic) at (0,1.0) {Domain logic};
  \node[comp, fill=blue!5]   (data)  at (0,0.0) {Data access};

  \node[comp, fill=green!10, minimum height=2.4cm] (down) at (4.5,0.8) {Downstream\\service};

  \draw[arrow] (data.east) -- (down.west |- data.east);

  \node[scope=9pt, draw=orange!80!black] (component_box) [fit=(api)(logic)(data)] {};
  \node[font=\small, text=orange!80!black, above=2pt of component_box.north] (component_text) {Component};

  \node[scope=7pt, draw=blue!70!black]  (s3) [fit=(component_box)(component_text)(down)] {};
  \node[font=\small, text=blue!70!black, below=2pt of s3.south] (integration_text) {Integration};

\end{tikzpicture}
  }
  \caption{Comparison between component and integration testing scopes.}
  \label{fig:testing_scope}
\end{figure}
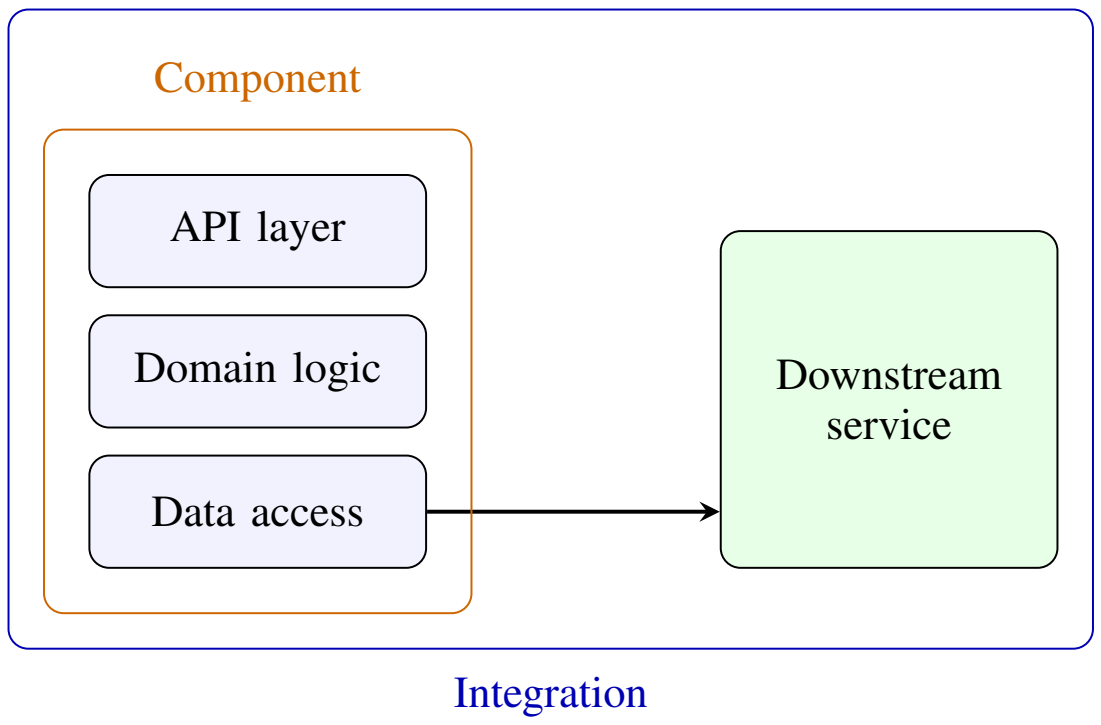

\subsection{Microservice Testing Levels}

Testing a microservice-based application introduces challenges not present in monolithic architectures: network calls add latency, while independent deployment means an update to an API can cause compatibility issues~\cite{GregorLena2025AToI,ponce2025107870}.
To manage this complexity, microservice testing is organized into multiple levels with distinct scopes and objectives~\cite{GregorLena2025BCaI}.
In this paper, we focus on the transition from component level to integration testing (see Figure~\ref{fig:testing_scope}).

\emph{Component} testing validates a single microservice through its exposed API while isolating external dependencies, such as downstream services and databases, typically using mocks or stubs~\cite{clemson2014testing}.
This isolation confines failures to the microservice under test and makes executions more deterministic, as the behavior of external dependencies is explicitly controlled.
However, this prevents tests from exercising actual inter-service interactions.
Component testing validates a microservice only against its own assumptions about how its collaborators behave rather than their actual runtime behavior.
Thus, failures that emerge only from interactions between services remain invisible at the component level.

\begin{figure}[t]
  \centering
  \resizebox{\columnwidth}{!}{
    \begin{tikzpicture}

  \node[pod, fill=blue!5,    minimum width=1.4cm, minimum height=0.85cm, font=\small] (test)   at (0,0)    {Client};
  \node[pod, fill=orange!10, minimum width=1.8cm, minimum height=0.85cm, font=\small] (target) at (2.8,0)  {Target\\service};
  \node[pod, fill=green!10,  minimum width=1.5cm, minimum height=0.85cm, font=\small] (a)      at (5.5,0)  {Service A};
  \node[pod, fill=red!15,    minimum width=1.5cm, minimum height=0.85cm, font=\small] (b)      at (8.0,0)  {Service B};

  \draw[arrow] (test) -- node[above=1pt, font=\footnotesize] {request} (target);
  \draw[arrow] (target) -- node[above=1pt, font=\footnotesize] {calls} (a);
  \draw[arrow] (a) -- node[above=1pt, font=\footnotesize] {calls} (b);

  \node[font=\footnotesize, text=red!70!black, align=center, below=0.15cm of b] {\texttt{500}\\downstream failure};

  \draw[dash_arrow, red!70!black] (b.north) to[bend right=36] (a.north);
  \draw[dash_arrow, red!70!black] (a.north) to[bend right=36] (target.north);
  \draw[dash_arrow, red!70!black] (target.north) to[bend right=36]
  node[above=2pt, font=\footnotesize, text=red!70!black] {failure propagates back} (test.north);

\end{tikzpicture}
  }
  \caption{Failure propagation across a chain of downstream microservice calls.}
  \label{fig:integration_call_graph}
\end{figure}
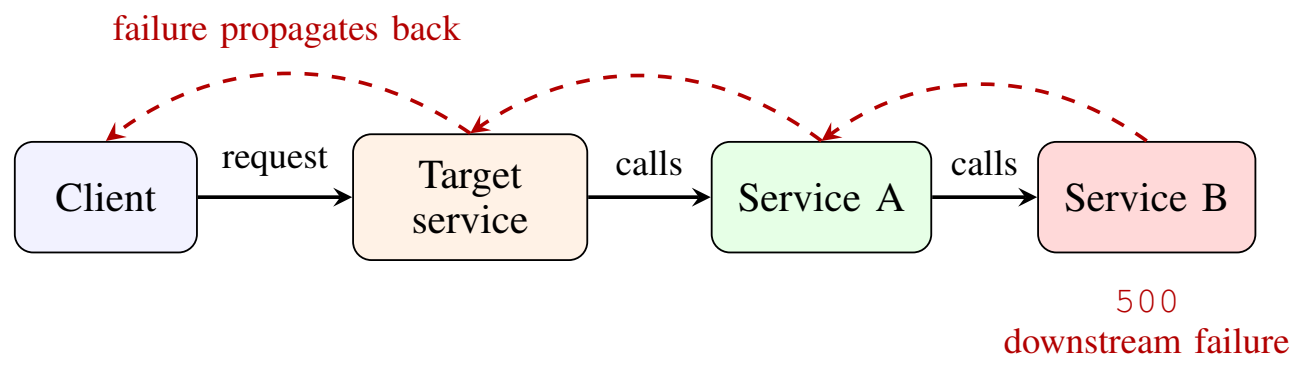

\emph{Integration} testing addresses this gap by deploying multiple interacting services and exercising them through their real interfaces~\cite{swebok_v4,GregorLena2025BCaI}.
In an integration test, requests are sent to a target service while its downstream services remain deployed and active.
An unexpected response from a downstream service may cause its caller to fail.
As illustrated in Figure~\ref{fig:integration_call_graph}, a failure originating in a downstream service can propagate back through multiple callers~\cite{SoldaniJacopo2025EMCF}, each of which may forward the response or replace it with a generic one.
Therefore, different failing executions may produce indistinguishable responses at the API boundary, hiding important differences in their internal behavior.

\subsection{Benchmarking Automated REST API Testing Tools: the \restgym Platform} \label{sec:background_limit}

Benchmarking infrastructures enable a fair, empirical comparison of automated testing tools under controlled and reproducible conditions, and are widely used in software engineering, for instance for software fuzzing~\cite{fuzzbench,unifuzz}, unit test generation~\cite{juge}, deep learning systems~\cite{giftbench}, and autonomous driving~\cite{biagiola2023sbft}.

\restgym~\cite{CorradiniDavide2025RAFI} is the state-of-the-art benchmarking infrastructure for automated black-box REST API testing.
To address the long-standing reproducibility challenges in REST API testing research, it provides a Docker-based~\cite{docker} environment pre-configured with a diverse suite of benchmark API services and testing tools.
The orchestration engine automates testing sessions across API services, performing multiple repetitions to account for non-deterministic behavior of test generators.
During each execution, \restgym collects the HTTP interactions produced by the testing tool via mitmproxy~\cite{mitmproxy}, together with coverage information from the service under test.
Following execution, \restgym processes the collected data to derive three main performance metrics:

\textbf{Operation Coverage} measures the number of API operations declared in the OpenAPI specification that are successfully exercised with a 2xx response.

\textbf{Code Coverage} measures the line, branch, and method coverage of the service under test, collected with the JaCoCo library~\cite{jacoco}.

\textbf{Unique Failures} estimates the number of distinct failures exposed during testing. Any request resulting in an HTTP 5xx response is flagged as a potential failure. Because automated testing tools generate thousands of requests, the same failure may be observed repeatedly, making raw failure counts unrepresentative. To report unique failures, \restgym implements a bucketing algorithm based on the Jaccard similarity index~\cite{Jaccard1901}.
Response bodies are normalized to eliminate dynamic values (e.g., timestamps and IDs) and tokenized.
Two failures are merged into the same bucket when their body similarity exceeds a threshold.

While this evaluation model is well suited to isolated services, applying it to distributed microservice systems exposes the following limitations:

\begin{itemize}
    \item \textit{Single-Service Assessment}: Existing benchmarking infrastructures, including \restgym, evaluate tools on single-service APIs, and therefore cannot capture failures that arise only from interactions with real downstream services.
    \item \textit{Insufficiency of Payload-Based Deduplication}: Bucketing 5xx responses by body similarity assumes that distinct failures produce distinguishable responses. In microservice systems, intermediate services or API gateways often turn downstream failures into the same generic error, so payload-only heuristics merge distinct failures and underestimate the number discovered.
    \item \textit{Lack of Failure Attribution}: Single-service benchmarks assume that all observed failures originate in the service under test. In microservice systems, an entry-point failure may instead have manifested in a downstream dependency and propagated through the call chain. Without benchmark-side observability of these internal interactions, the evaluator cannot distinguish the two.
    \item \textit{Random Seed Control}: Testing tools commonly rely on pseudo-random generators, so without systematic seed control by the orchestrator, executions are hard to reproduce and cross-tool comparisons are less controlled.
\end{itemize}

\section{\restberus}

\restberus is a Kubernetes-based~\cite{kubernetes} benchmarking infrastructure designed to evaluate automated REST API testing tools against microservice architectures.
Kubernetes provides the orchestration mechanisms needed to deploy complete microservice dependency graphs, isolate concurrent experimental runs, and consistently manage the lifecycle and instrumentation of the services involved in each evaluation.
Extending benchmarking REST API testing tools from isolated services to distributed systems, \restberus automates the deployment of microservice architectures and collects runtime telemetry across service boundaries.
This enables the evaluation of black-box testing tools in integration testing scenarios.

The design of RESTberus is guided by three main goals derived from the limitations discussed in Section~\ref{sec:background_limit}.
First, \textit{distributed deployment} enables testing tools to exercise real inter-service interactions rather than isolated services with mocked dependencies.
Second, \textit{internal observability} makes the execution of requests across service boundaries visible to \restberus, enabling the attribution of observed failures to target or downstream services and trace-aware failure deduplication.
Third, \textit{controlled execution} isolates individual runs, standardizes the deployment and execution of testing tools and benchmark systems, and manages random seeds to support reproducible comparisons across experiments.

\subsection{Architecture}
\begin{figure}[t!]
  \centering
  \resizebox{\columnwidth}{!}{
    \begin{tikzpicture}

  \node[pod, fill=yellow!15, text width=1.7cm, minimum height=0.75cm, font=\small] (sidecar) at (1.0, 2.7) {mitmproxy};
  \node[pod, fill=orange!15, text width=1.7cm, minimum height=0.75cm, font=\small] (api)     at (1.0, 0.9) {API Target};

  \node[pod, fill=red!15,    text width=2.0cm, minimum height=0.75cm, font=\small] (tool)    at (8.0, 2.7) {Testing Tool};

  \node[pod, fill=green!15,  text width=1.4cm, minimum height=0.75cm, font=\small] (envoy)   at (4.6, 0.9) {Envoy\\proxy};
  \node[pod, fill=green!15,  text width=1.8cm, minimum height=0.75cm, font=\small] (down)    at (8.0, 0.9) {Downstream\\Service};

  \node[draw, dotted, thick, blue!90, rounded corners, inner sep=0.32cm, fit=(api) (sidecar)] (target_api) {};
  \node[blue!90, font=\footnotesize, above=2pt of target_api.north] (target_label) {Target Service (API Pod)};

  \node[draw, dotted, thick, blue!70, rounded corners, inner sep=0.25cm, fit=(tool)] (tool_pod) {};
  \node[blue!70, font=\footnotesize, above=2pt of tool_pod.north] (tool_label) {Tool Pod};

  \node[draw, dotted, thick, gray!80!black, rounded corners, inner sep=0.25cm, fit=(down) (envoy)] (downstream_pod) {};
  \node[gray!80!black, font=\footnotesize, below=2pt of downstream_pod.south] (down_label) {Downstream Services (Other API Pods)};

  \node[draw, dashed, black!60, rounded corners, inner sep=0.35cm, fit=(target_api) (target_label) (tool_pod) (tool_label) (downstream_pod) (down_label)] (ns) {};
  \node[black!80, font=\small, above=2pt of ns.north] {Kubernetes Namespace};

  \draw[arrow] (tool.west) -- node[above=1pt, font=\footnotesize] {Send requests} (sidecar.east);
  \draw[arrow] (sidecar.south) -- node[right=2pt, font=\footnotesize] {Forward} (api.north);
  \draw[arrow, dotted] (api.east) -- node[above=2pt, font=\footnotesize, pos=0.35] {Call} (envoy.west);
  \draw[arrow] (envoy.east) -- node[above=1pt, font=\footnotesize] {Forward} (down.west);

\end{tikzpicture}
  }
  \caption{Runtime topology of a RESTberus run within a Kubernetes namespace.}
  \label{fig:namespace_topology}
\end{figure}
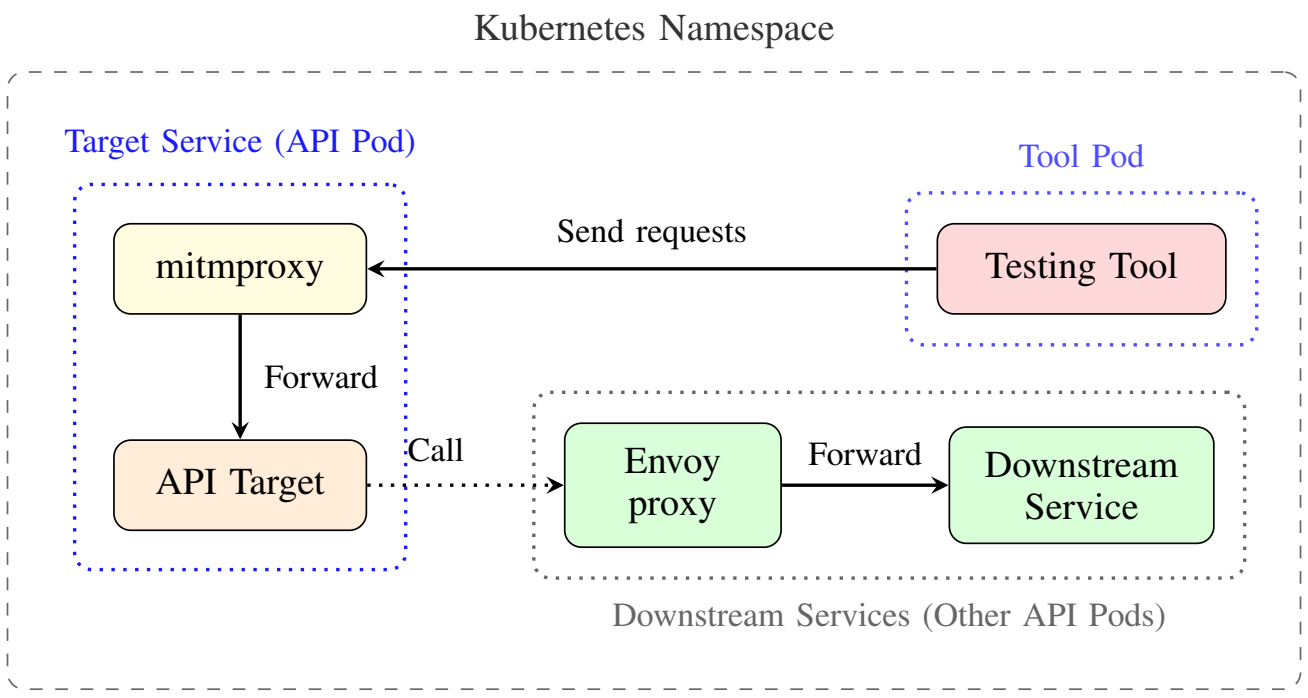

\restberus adopts a client-server architecture.
An orchestrator engine coordinates the lifecycle of experiments and persists runtime telemetry.
Users interact with the platform through a command-line interface to submit experiment configurations, monitor execution progress, and export evaluation reports.
This decoupling allows the orchestrator to run either locally or directly inside a Kubernetes cluster, while benchmark systems and testing tools are instantiated as containerized workloads in Kubernetes.
For each run, RESTberus deploys the target microservice together with its downstream services in an isolated namespace and launches the testing tool as a separate Kubernetes Job.

Figure~\ref{fig:namespace_topology} illustrates this runtime topology of a RESTberus run.
Each component (the testing tool, the target service, and each downstream service) executes inside its own \emph{pod}, the basic Kubernetes unit grouping one or more tightly coupled containers.
HTTP requests sent to the target service pass through a \texttt{mitmproxy} sidecar, which records request and response data observed at the API entry point.
To complement these API-boundary observations, RESTberus instruments the deployed downstream services with Istio/Envoy proxies, which observe inter-service HTTP calls and generate distributed traces describing how each request propagates across the microservice chain.
Test traffic reaches the target through mitmproxy without passing through Envoy, so client-to-target interactions are recorded at the API boundary, while distributed traces capture downstream inter-service calls.
A trace is composed of individual \emph{spans}, each representing one service-level operation or network call along that execution path.
RESTberus also records code coverage measurements for the exercised services.
The collected HTTP interactions, trace spans, and coverage measurements are stored in a local SQLite database~\cite{sqlite} and later processed to compute aggregate evaluation metrics and reports.

\subsection{Experiment Orchestration and Run Lifecycle}

In \restberus, evaluations are organized into \textit{experiments} and \textit{runs}.
An experiment specifies a set of testing tools, a set of benchmark systems, and a number of repetitions.
Each run is one execution of a single testing tool against a dedicated instance of a benchmark system.
The orchestrator schedules these runs and manages their complete lifecycle.

The lifecycle of each evaluation run proceeds through three sequential phases: \emph{deployment}, \emph{execution}, and \emph{cleanup}.
During deployment, the orchestrator creates an ephemeral Kubernetes namespace dedicated to the specific run and deploys the benchmark system using Helm.
The deployment includes the target API service alongside its downstream services and any required databases.
Assigning each run to a separate namespace provides logical isolation between concurrent runs, preventing interference between their Kubernetes resources and service names.
The namespace is also configured with the instrumentation required by \restberus, including Istio/Envoy proxies for the downstream services and a \texttt{mitmproxy} sidecar for the target service.

Before executing the testing tool, \restberus verifies that the deployed system is ready.
It waits until all pod containers report a ready state and then probes the target API until it responds successfully.
If either condition is not satisfied within a configured timeout, deployment is aborted, the namespace is marked as failed, and the run is scheduled for retry.

Once the deployment satisfies all readiness checks, the run enters the execution phase.
The orchestrator launches the testing tool as a Kubernetes Job inside the same namespace, allowing it to communicate directly with the target service.
To accommodate tools with different command-line interfaces and configuration formats without modifying tool source code, each tool is packaged in a container image with a standardized launch script.
The orchestrator mounts the OpenAPI specification and input dictionaries required by the tool into the tool container, and passes runtime parameters (e.g., target URL and interaction budget) as environment variables.
To control randomness in test generation, the orchestrator computes a distinct pseudo-random seed for each run based on the experiment seed and repetition index, injecting it into the tool container.
Crucially, the tool pod is deliberately deployed without the Envoy sidecar normally injected by Istio, as shown in Figure~\ref{fig:namespace_topology}.
Since Envoy enforces strict HTTP validation, routing test traffic through it could reject malformed requests before they reach the target.
Disabling this sidecar preserves the original traffic generated by the testing tool.
During execution, background workers collect HTTP interactions, trace spans, and code coverage.

Following execution, the run concludes with the cleanup phase, regardless of whether testing completed successfully or aborted due to an error.
Before tearing down the deployment, the orchestrator performs a final telemetry collection to capture observations from the last execution interval.
It then deletes the namespace, releasing cluster resources for subsequent runs.
All collected telemetry is stored for trace reconstruction and failure analysis, as described in Sections~\ref{sec:reconstruction} and~\ref{sec:failure-analysis}.

\subsection{Telemetry and Trace Reconstruction} \label{sec:reconstruction}

To reconstruct how a test request propagates across microservices, \restberus must associate each HTTP interaction recorded at the API entry point with the distributed trace describing its downstream calls.
To establish this association, the \texttt{mitmproxy} sidecar injects a unique request identifier into each request entering the target service.
This identifier is propagated across inter-service calls and recorded by Jaeger as an attribute of the corresponding trace spans.
\restberus uses this identifier to retrieve the spans associated with each HTTP interaction.

Once the spans for a request are retrieved, \restberus reconstructs the observed inter-service call paths.
Spans are ordered by their start times to establish their chronological order, while the parent link of each span is used to trace its call path back to the target service.
These reconstructed paths provide the execution context for the failure analysis described in Section~\ref{sec:failure-analysis}.

\subsection{Trace-Aware Failure Analysis} \label{sec:failure-analysis}

\restberus analyzes observed failures through two complementary tasks.
\emph{Failure attribution} uses the reconstructed traces to classify each failure as a target failure, when it manifests in the target service, or a downstream failure, when it manifests in one of its dependencies.
\emph{Failure deduplication} groups equivalent failure occurrences into failure buckets to estimate the number of distinct failures.

Failure attribution identifies where a failure manifested in the observed trace.
\restberus examines the 5xx spans associated with a failing HTTP interaction and selects the latest-starting one, since downstream failures may propagate through multiple callers.
A failure is attributed to a downstream service when this span belongs to a service other than the target, and to the target service otherwise.
If no 5xx span is available for an HTTP interaction, \restberus classifies the failure as target-side by default.
Relying on start time rather than parent links allows this heuristic to operate even when intermediate spans are missing, e.g., when a downstream call fails before the callee records its span or when a service times out leaving a gap in the captured execution path.

Failure deduplication decides whether two observed failures are the same or different.
\restberus adopts two matching criteria from prior work and introduces a third based on distributed traces.
The \emph{payload} criterion reproduces \restgym's response body similarity bucketing described in Section~\ref{sec:background_limit}, with the same threshold of $0.7$~\cite{CorradiniDavide2025RAFI}, retrieved from its replication package.
The \emph{operation} criterion, adopted from MacroHive~\cite{GiamatteiLuca2024AFaR}, matches failures associated with the same API operation, regardless of their response bodies.
The \emph{trace} criterion matches failures whose latest-starting 5xx spans belong to the same service, or whose reconstructed call paths are compatible.
Compatibility requires the set of service-operation pairs from one path to be a subset of the other, which allows partial traces with missing spans to merge with more complete ones.

Each criterion has limitations.
The payload criterion merges distinct failures masked by a generic error body, as discussed in Section~\ref{sec:background_limit}.
The operation criterion merges distinct failures reached through the same operation, as happens when an operation calls several downstream services.
The proposed trace criterion requires spans, which may be unavailable for every failure, since a failure raised by the target service without any downstream call produces no span.

Based on these criteria, \restberus implements five deduplication strategies.
The \texttt{payload\_only} and \texttt{operation\_only} strategies apply their respective criteria independently and serve as baselines.
\texttt{trace\_payload} merges two traced failures only when both the trace and payload criteria hold, and applies the payload criterion alone to untraced failures.
Similarly, \texttt{trace\_operation} combines trace compatibility with operation matching for traced failures and applies operation matching to untraced failures.
The \texttt{hybrid} strategy handles traced failures as \texttt{trace\_payload}, but groups untraced failures by API operation.
This hybrid design exploits complementary evidence depending on trace availability: for traced failures, payload similarity helps distinguish different failure manifestations along compatible execution paths; for untraced failures, operation-based grouping avoids merging failures from different API operations solely because they produce the same generic response.
The three trace-aware strategies never merge traced and untraced failures, since execution-trace evidence is available for only one failure and cannot support a direct comparison of their internal behavior.

\section{Empirical Setup}

\subsection{Research Questions}

\textbf{RQ1 (Failure Deduplication)}:
\textit{To what extent do trace-aware strategies improve failure deduplication accuracy in microservice systems?}

Existing approaches estimate unique failures using different criteria: \restgym groups HTTP 5xx responses by body similarity, while MacroHive counts distinct failing API operations.
In microservice architectures, however, different downstream failures may produce indistinguishable responses or occur through the same API operation, causing these deduplication criteria to merge distinct failures.
RQ1 investigates whether trace-aware strategies improve deduplication accuracy over these existing approaches.
We evaluate the strategies against human-calibrated LLM reference labels for sampled failure pairs, and assess the effectiveness of existing testing tools according to the best-performing strategy.

\textbf{RQ2 (Downstream Failure Attribution)}: \textit{What proportion of unique failures observed during REST API testing are attributed to downstream dependencies rather than to the target service?}

In microservice architectures, an HTTP 5xx response observed at the entry point may reflect a failure manifested in a downstream microservice rather than in the target service.
Since entry-point observations alone cannot distinguish these cases, RQ2 uses \restberus trace-based attribution to quantify the prevalence of unique failures attributed to downstream services.

\textbf{RQ3 (Platform Equivalence)}: \textit{Does \restberus produce equivalent evaluation results to \restgym when testing single-service applications?}

Although \restberus extends benchmarking to microservice systems, its Kubernetes-based architecture introduces differences in orchestration and networking.
To assess whether these changes influence evaluation outcomes, RQ3 compares \restberus and \restgym using the same testing tools and a common subset of single-service benchmarks under the same experimental conditions.
We assess whether differences in the resulting evaluation metrics remain within predefined margins of practical equivalence.

\subsection{Testing Tools}

To evaluate \restberus across representative testing methodologies, we considered the automated black-box REST API testing tools already assessed in prior benchmarking studies~\cite{CorradiniDavide2025RAFI, pasqua2026sbft}.
We selected tools from this pool according to the following criteria:
\begin{enumerate}
    \item \textit{OAS 3.0 Compatibility}: Tools must support OpenAPI 3.0 specifications and generate tests for the benchmark APIs;
    \item \textit{Availability}: Tools must have publicly available implementations;
    \item \textit{Technological Diversity}: The tool set should cover different test generation approaches to provide a comprehensive assessment of REST API testing techniques.
\end{enumerate}

Based on these criteria, we included six tools:

\begin{itemize}
    \item \textbf{\deeprest}~\cite{deeprest}:
    Uses reinforcement learning to handle implicit operation dependencies and parameter constraints;

    \item \textbf{\morest}~\cite{morest}:
    Constructs a graph from the OpenAPI specification to model operations, data schemas, and producer-consumer relationships;

    \item \textbf{\restest}~\cite{restest}:
    Applies constraint-based testing to handle inter-parameter dependencies, including mutually exclusive or conditionally required parameters;

    \item \textbf{\restler}~\cite{restler}:
    Uses grammar-based fuzzing, compiling OpenAPI specifications in request generation grammars;

    \item \textbf{\resttestgen}~\cite{resttestgen}:
    Constructs a graph from the OpenAPI specification based on input-output dependencies between API operations to guide test generation;

    \item \textbf{\schemathesis}~\cite{schemathesis}:
    Applies property-based testing to generate inputs from OpenAPI specifications.
\end{itemize}

All six tools exercise the target service through its REST API, using its OpenAPI specification to guide test generation, and observe the responses at that boundary.
Since internal service interactions and distributed traces are not exposed to the tools, their test generation logic remains unchanged when moving from isolated services to microservice deployments.

\subsection{Experimental Objects}

To evaluate \restberus across both single-service baselines and distributed environments, we selected three benchmark suites:

\textbf{Train Ticket}~\cite{ZhouXiang2021FAaD}: A train booking system composed of 41 microservices, recognized by a recent systematic review as the de-facto standard benchmark for large microservice architectures~\cite{ponce2025107870}, and used in prior work on integration testing and fault localization~\cite{GregorLena2025BCaI,ibm_cloud22}.
Its distributed architecture and inter-service dependencies make it suitable for evaluating how REST API testing tools exercise multi-service interactions and expose downstream failures.
Following the integration testing configurations by Gregor et al.~\cite{GregorLena2025BCaI}, we organized Train Ticket into 19 test subsets, each comprising one target service and all the downstream services it directly or transitively invokes.
This allows tools to exercise real inter-service interactions without deploying the complete application for every experiment.
To support controlled testing and failure measurement, we applied the following adaptations:

\begin{itemize}
    \item \textit{Circuit-Breaker Fallback Removal}: In the original Train Ticket implementation, operations are instrumented with Hystrix~\cite{hystrix} circuit breakers. When a downstream call fails, the circuit breaker intercepts the exception and returns a synthetic response (e.g., an empty payload or an HTTP 200 OK response).
    Although these fallbacks improve service resilience in production, they may interfere with the failure oracle of automated black-box REST API testing tools, which detect failures through HTTP 5xx responses.
    A downstream failure converted into a successful response is never reported, since the tools only see the response returned at the API boundary.
    Thus, we disabled these fallback handlers so that otherwise unhandled downstream failures can propagate back to the caller as HTTP 5xx responses and remain observable.

    \item \textit{Timeout Reduction}: We reduced the socket timeout of inter-service HTTP calls from 5 seconds to 500\,ms, so that boundary payloads that make downstream services hang do not stall the testing tool and lower test throughput.

    \item \textit{Distributed Tracing}: We integrated Spring Cloud Sleuth~\cite{spring_cloud_sleuth} to enable trace collection across services, by propagating W3C trace context and the injected request identifier across outgoing HTTP calls.

    \item \textit{OpenAPI Specification Migration}: We migrated legacy Swagger 2.0 specifications to OpenAPI 3.0 using springdoc~\cite{springdoc} and validated all resulting specifications with Spectral~\cite{spectral} and manual review.

    \item \textit{Coverage and Telemetry Instrumentation}: We configured services to retain exception details in HTTP error responses, preserving diagnostic information for response-based failure analysis.
    We also packaged each container image with the JaCoCo agent to collect runtime code coverage.

\end{itemize}

\textbf{PetClinic}~\cite{spring_petclinic}: A distributed implementation of the Spring PetClinic application composed of 7 microservices.
As a smaller distributed benchmark, it complements Train Ticket and allows us to assess trace-aware failure analysis across different microservice applications and dependency structures.
We applied the following adaptations:
\begin{itemize}
    \item \textit{Removal of External Dependencies and Static Routes:} We removed dependencies on services outside the benchmark deployment and excluded static-resource routes, so that observed failures come from REST operations within the deployed system.

    \item \textit{OpenAPI 3.0 Specification Generation:} We integrated springdoc~\cite{springdoc} to generate OpenAPI 3.0 specifications for the gateway-exposed REST routes.

    \item \textit{Distributed Tracing and Telemetry Instrumentation:} Unlike Train Ticket, PetClinic natively supports distributed tracing. We configured it to sample 100\% of traces.
    As with Train Ticket, we configured error handling to retain exception details in HTTP error responses and instrumented the services with JaCoCo to collect runtime code coverage.
\end{itemize}

\textbf{Single-Service Benchmarks}: To address RQ3, which evaluates whether \restberus reproduces \restgym results on single-service applications, we included 9 of the 17 services available in the \restgym repository~\cite{restgym_repo}.
We selected the services on which all six evaluated tools could execute to completion on both platforms, ensuring a balanced paired design across all evaluation metrics.

\subsection{Experimental Protocol}

To ensure fair comparison across tools with different generation speeds, we bounded each run by an interaction budget of 10,000 HTTP requests rather than a wall-clock timeout.
A preliminary convergence analysis showed that growth slows well before this limit. Code coverage and unique failures reached 95\% of their final values within 5,500 and 6,000 requests, respectively, while all metrics reached 99\% by 9,500 requests.
Each tool-service combination was executed ten times, with a distinct random seed to account for non-deterministic exploration~\cite{ArcuriAndrea2011Apgf} while supporting reproducibility.

For RQ3, we evaluated the same six tools on nine single-service benchmarks using both \restberus and \restgym, with ten repetitions and the same 10,000-request budget on each platform.
For each tool-service combination, we averaged the ten repetitions separately for each platform, obtaining a matched pair of measurements for the statistical analyses.

\restberus performs an eligibility check to verify run validity, thus ensuring data quality: a run is deemed valid only if it completed its full interaction budget and recorded coverage samples throughout execution.
Runs that terminate prematurely or stall are automatically retried with the same random seed and parameters.
Ten runs required a retry due to temporary cluster resource exhaustion.

\subsection{Evaluation Metrics}

Runs are processed to compute metrics to answer each research question.
We consider HTTP 5xx responses observed at the API entry point as potential failures, consistently with \restgym.
To estimate the number of \textit{Unique HTTP 5xx Failures}, we group these observations into buckets using the deduplication strategies described in Section~\ref{sec:failure-analysis}.
Each bucket represents an estimated distinct failure manifestation.

In RQ1, we assess deduplication against reference labels using \textit{Accuracy} (the proportion of correctly classified failure pairs), \textit{Precision} (the proportion of predicted duplicates that are actual duplicates), and the \textit{Matthews Correlation Coefficient (MCC)}~\cite{ChiccoDavide2020Taot} (the correlation between predicted and reference binary labels, i.e., \textsc{Duplicate} or \textsc{Distinct}).
MCC accounts for all four outcomes of binary classification and is informative under class imbalance, ranging from $-1$ (perfect disagreement) through $0$ (no correlation) to $+1$ (perfect agreement).
To evaluate testing tool effectiveness, we measure \textit{Operation Coverage}, \textit{Line Coverage}, and \textit{Unique HTTP 5xx Failures}.

In RQ2, following Section~\ref{sec:failure-analysis}, we partition unique failures into \textit{Target Failures} and \textit{Downstream Failures}.
We report both counts, their sum (\textit{Total Failures}), and the \textit{Downstream Failure Ratio} (i.e., the percentage of unique failures attributed to downstream dependencies).

In RQ3, we evaluate five metrics across single-service benchmarks: \textit{Operation Coverage}, \textit{Line Coverage}, \textit{Branch Coverage}, \textit{Method Coverage}, and \textit{Unique HTTP 5xx Failures}.

\subsection{Experimental Procedure}

All experiments were executed on an isolated Linux virtual machine equipped with 64 vCPUs of an Intel Xeon 6342 processor and 96\,GB of RAM.
The Kubernetes cluster environment was managed using k3s~\cite{k3s}.
Executing the full experimental suite across Train Ticket, PetClinic, and the single-service benchmarks required approximately 10 days.

To establish the reference labels for RQ1, we collected an evaluation dataset of failure pairs using a two-tier stratified sampling strategy across all experimental runs.
For each run, we sampled 15 failure pairs uniformly at random to capture general testing behavior, and 5 additional pairs where both failures carried distributed traces, ensuring representation of downstream microservice interactions.
As some runs did not trigger enough failures, fewer pairs were sampled in those runs, yielding $N = 21{,}997$ pairs in total.
Each pair was classified as \textsc{Duplicate} or \textsc{Distinct} by a multi-model LLM judge ensemble (Qwen 3.8 27B and Gemma 4 31B, with DeepSeek V4 Flash resolving ties).
The judges used entry-point HTTP responses, trace call sequences, and container logs, with four few-shot demonstration examples and temperature set to zero to minimize sampling variability.
To confirm the reliability of the LLM reference labels, we manually annotated a calibration sample of 346 failure pairs and measured agreement using Cohen's $\kappa$~\cite{Cohen1960}.
The LLM consensus achieved 87.9\% agreement with these annotations (Cohen's $\kappa = 0.758$), indicating substantial agreement~\cite{Richard1977TMoO}.
The two primary judges agreed on 89.0\% of pairs ($\kappa = 0.780$), with the arbiter resolving the remaining 11.0\% of disagreements.

We evaluated the five deduplication strategies against both the LLM reference labels and the human annotations, using McNemar's test~\cite{McNemar1947} to assess differences in classification accuracy.
We then compared testing tool effectiveness across Train Ticket and PetClinic using the best-performing deduplication strategy.

In RQ2, we investigated failure attribution by categorizing unique failures into target and downstream failures according to the attribution criterion of Section~\ref{sec:failure-analysis}.
In particular, we computed unique-failure counts using the best-performing deduplication strategy identified in RQ1.
We compared tools on target, downstream, and total unique failure counts across benchmark services and computed the proportion of failures attributed to downstream dependencies.

In RQ3, we assessed platform equivalence between \restberus and \restgym across the nine single-service benchmarks.
To ensure a like-for-like comparison, unique failures were computed using the \texttt{payload\_only} strategy on both platforms, reproducing the response-based bucketing adopted by \restgym.
In fact, as the single-service benchmarks are not traced, \texttt{trace\_payload} falls back to the same payload comparison and would yield identical counts, whereas \texttt{trace\_operation} and \texttt{hybrid} fall back to grouping failures by operation instead of by response body and would yield different counts by design.
We assessed systematic differences in the five evaluation metrics and tested practical equivalence against predefined margins of $\pm 1$ operation, $\pm 1$ percentage point for code coverage, and $\pm 1$ unique failure.
Equivalence was established when the Hodges--Lehmann 90\% confidence interval~\cite{HodgesLehmann1963} for a paired difference lay entirely within the corresponding margin, according to equivalence testing principles~\cite{Wellek2010}.

Comparisons of testing-tool effectiveness in RQ1 and RQ2, as well as comparisons between platforms in RQ3, used two-sided paired Wilcoxon signed-rank tests~\cite{Wilcoxon1945} on the means of the ten repetitions, with a significance threshold of $\alpha = 0.05$.

\section{Results}

\begin{table}[t]
  \centering
  \footnotesize
  \caption{RQ1: Failure deduplication performance across evaluation sets. Best results in bold; underlined strategies are not statistically different from the best.}
  \label{tab:rq1_dedup_validity}
  \setlength{\tabcolsep}{6pt}
  \begin{tabular}{@{} l c c c @{}}
    \toprule
    Heuristic & Accuracy (\%) & Precision (\%) & MCC \\
    \midrule
    \multicolumn{4}{@{}l}{\textit{Traced (LLM consensus, $N = 4{,}725$)}} \\
    \midrule
    \texttt{payload\_only}    &            82.0 &            80.6 &           0.633 \\
    \texttt{operation\_only}  &            64.6 &            64.8 &           0.264 \\
    \texttt{trace\_payload}   &   \textbf{91.0} &   \textbf{91.7} &  \textbf{0.818} \\
    \texttt{trace\_operation} &            77.8 &            76.6 &           0.547 \\
    \texttt{hybrid}           &   \textbf{91.0} &   \textbf{91.7} &  \textbf{0.818} \\
    \midrule
    \multicolumn{4}{@{}l}{\textit{All Pairs (LLM consensus, $N = 21{,}997$)}} \\
    \midrule
    \texttt{payload\_only}    &            74.1 &            69.5 &           0.498 \\
    \texttt{operation\_only}  &            77.4 &            75.1 &           0.550 \\
    \texttt{trace\_payload}   &            78.5 &            74.0 &           0.582 \\
    \texttt{trace\_operation} &            83.1 &            84.4 &           0.662 \\
    \texttt{hybrid}           &   \textbf{87.1} &   \textbf{90.2} &  \textbf{0.745} \\
    \midrule
    \multicolumn{4}{@{}l}{\textit{Human calibration sample ($N = 346$)}} \\
    \midrule
    \texttt{payload\_only}    &            77.5 &            70.0 &           0.584 \\
    \texttt{operation\_only}  &            79.8 &            74.6 &           0.606 \\
    \texttt{trace\_payload}   &            80.6 &            73.4 &           0.640 \\
    \texttt{trace\_operation} & \underline{90.5} & \underline{91.4} & \underline{0.809} \\
    \texttt{hybrid}           &   \textbf{92.2} &   \textbf{94.3} &  \textbf{0.845} \\
    \bottomrule
  \end{tabular}
\end{table}

\subsection{RQ1 (Failure Deduplication)}

Table~\ref{tab:rq1_dedup_validity} reports the performance of the five deduplication strategies.
On traced pairs, \texttt{hybrid} and \texttt{trace\_payload} achieve identical results, since they apply the same matching criteria when traces are available.
Both outperform \texttt{payload\_only} (91.0\% vs. 82.0\% accuracy), because traces separate distinct downstream failures that gateways mask behind the same error body.
\texttt{trace\_operation} drops to 77.8\% accuracy on these pairs.
In more than half of the traced pairs, both failures manifest in the target after successful downstream calls, so the trace adds no evidence and only the response body separates distinct failures of the same operation, e.g., a timeout from a null pointer exception.
When both failures manifest in the same downstream service, the entry operation is also a poor proxy, as the same failure is often reached through different operations.
These limitations also explain why \texttt{operation\_only} achieves the lowest accuracy on these pairs (64.6\%).
Across all sampled pairs, \texttt{hybrid} achieves the best results (87.1\% accuracy, 0.745 MCC).
McNemar's test confirms that, on these pairs, \texttt{hybrid} differs significantly from every other strategy ($p < 0.001$).

To mitigate potential evaluation bias arising from using LLM consensus, we evaluated all five strategies on the manually annotated calibration sample.
On this manual benchmark, \texttt{hybrid} again achieves the highest accuracy (92.2\%) and MCC (0.845).
\texttt{hybrid} significantly outperforms \texttt{payload\_only}, \texttt{operation\_only}, and \texttt{trace\_payload} ($p < 0.001$), while \texttt{trace\_operation} achieves comparably high accuracy, with no statistically significant difference from \texttt{hybrid} ($p = 0.109$) and only 10 disagreements across 346 pairs.
Thus, \texttt{hybrid} is the most accurate deduplication strategy whether the reference labels come from the LLM judges or from manual annotation.

\begin{table}[htbp]
  \centering
  \footnotesize
  \caption{RQ1: Per-tool effectiveness on Train Ticket and PetClinic, averaged across services. Best results in bold; underlined values are not statistically different from the best.}
  \label{tab:rq1_total}
  \setlength{\tabcolsep}{3.5pt}
  \begin{tabular}{@{} l r r r @{}}
    \toprule
    Tool         & {\shortstack{Operation\\Coverage}} & {\shortstack{Line\\Coverage (\%)}} & {\shortstack{Unique 5xx\\failures}} \\
    \midrule
    \deeprest    & \textbf{11.59}   & \textbf{55.1}    & 9.80            \\
    \morest      & 5.83             & 45.5             & 5.45            \\
    \restest     & 7.98             & 51.4             & 9.21            \\
    \restler     & 7.09             & 50.0             & 7.74            \\
    \resttestgen & 10.21            & \underline{52.8} & 8.98            \\
    \schemathesis & 8.15             & 49.4             & \textbf{14.56}  \\
    \midrule
    Average      & 8.48             & 50.7             & 9.29            \\
    \bottomrule
  \end{tabular}
\end{table}

Therefore, we use \texttt{hybrid} to assess the unique failures discovered by each tool, alongside coverage metrics.
Table~\ref{tab:rq1_total} reports these metrics averaged across repetitions and benchmark services in Train Ticket and PetClinic.

\deeprest\ achieves the highest operation coverage (11.59), and ties statistically with \resttestgen\ for highest line coverage (55.1\% vs. 52.8\%, \(p = 0.083\)).
\schemathesis\ discovers the highest number of unique failures (14.56 per run), significantly outperforming all other tools ($p < 0.001$).

\begin{resultbox}
RQ1: The proposed \texttt{hybrid} strategy outperforms all alternative deduplication strategies, achieving 87.1\% accuracy and 90.2\% precision across an evaluation sample including both traced and untraced failures, with improvements of up to 13.0 and 20.7 percentage points, respectively.
Its effectiveness is further supported by the highest observed accuracy (92.2\%) on human-annotated failure pairs.
\end{resultbox}

\subsection{RQ2 (Downstream Failure Attribution)}

To answer RQ2, we investigate how frequently testing tools trigger failures beyond the target service.
Using the latest-starting 5xx trace span, we classify failures as either local to the target service or manifesting in a downstream service.

Table~\ref{tab:rq2_total} reports the cumulative number of unique target and downstream failures detected by each tool across Train Ticket and PetClinic, averaged over repetitions per service and summed across services.

\begin{table}[b]
  \centering
  \footnotesize
  \caption{RQ2: Target and downstream failure attribution on Train Ticket and PetClinic. Best results in bold; underlined values are not statistically different from the best.}
  \label{tab:rq2_total}
  \setlength{\tabcolsep}{3.5pt}
  \begin{tabular}{@{} l r r r r @{}}
    \toprule
    Tool & {Target} & {Downstream} & {Total} & {\shortstack{Downstream\\ratio (\%)}} \\
    \midrule
    \deeprest    & 167.2 & \textbf{28.8} & 196.0 & 14.7 \\
    \morest      & 96.8 & 12.1 & 108.9 & 11.1 \\
    \restest     & 158.2 & \underline{26.1} & 184.3 & 14.2 \\
    \restler     & 139.8 & 14.9 & 154.7 & 9.6 \\
    \resttestgen & 154.7 & \underline{25.0} & 179.7 & 13.9 \\
    \schemathesis & \textbf{267.6} & \underline{23.7} & \textbf{291.3} & 8.1 \\
    \midrule
    Average      & 164.0 &   21.8 & 185.8 & 11.7 \\
    \bottomrule
  \end{tabular}
\end{table}

All six tools expose downstream failures, which account for an average of 11.7\% of their unique failures, ranging from 8.1\% to 14.7\% across tools.
\deeprest\ detects the most downstream failures (28.8), although its differences from \restest, \resttestgen, and \schemathesis\ are not statistically significant.
Conversely, \schemathesis\ detects the most target-service failures (267.6), significantly outperforming all other tools.
These results show that tool rankings depend on where failures manifest, highlighting information that cannot be obtained from API-boundary responses alone.

\begin{resultbox}
RQ2: Downstream failures account for an average of 11.7\% of unique failures across all six testing tools.
By distinguishing these failures from target-service failures, \restberus trace-based attribution provides an additional dimension for assessing testing-tool effectiveness.
\end{resultbox}

\subsection{RQ3 (Platform Equivalence)}

To answer RQ3, we investigate whether \restberus reproduces the evaluation results of \restgym on single-service benchmarks, despite differences in orchestration and networking.
We compare the platforms across five metrics: operation coverage, line, branch, and method coverage, and unique HTTP 5xx failures.
We report unique failures using \texttt{payload\_only} on both platforms, to reproduce \restgym's original deduplication criterion.
For each tool-service combination, the mean over ten repetitions on each platform forms a matched pair for statistical analysis.

Table~\ref{tab:rq3_aggregate} reports the aggregate mean performance of both platforms.
To compare the platforms, we evaluate two-sided Wilcoxon signed-rank tests to check for systematic bias, and Hodges-Lehmann 90\% confidence intervals against predefined practical equivalence margins.
These margins represent the smallest differences of practical interest: \(\pm 1\) operation (the minimal discrete unit), \(\pm 1\) percentage point for code coverage metrics (comparable to the variation between repeated runs of the same tool), and \(\pm 1\) for unique failures.

For all five metrics, the two-sided paired Wilcoxon signed-rank tests confirm no statistically significant differences between platforms (\(p \ge 0.21\)).
Their Hodges-Lehmann 90\% confidence intervals also lie entirely within the equivalence margins.
These results establish practical equivalence between the two platforms under the evaluated conditions.

\begin{table}[t]
  \centering
  \footnotesize
    \caption{RQ3: Comparison \restberus vs \restgym across benchmark services. Underlined values denote statistical equivalence.}
  \label{tab:rq3_aggregate}
  \setlength{\tabcolsep}{4pt}
  \begin{tabular}{@{}lrr@{}}
    \toprule
    Metric & \restgym & \restberus \\
    \midrule
    Operation coverage    & 11.12 & \underline{11.51} \\
    Line coverage (\%)    &  56.2 &  \underline{56.7} \\
    Branch coverage (\%)  &  23.9 &  \underline{23.8} \\
    Method coverage (\%)  &  59.4 &  \underline{59.9} \\
    Unique 5xx failures   &  7.42 &  \underline{7.14} \\
    \bottomrule
  \end{tabular}
\end{table}

\begin{resultbox}
RQ3: Across the considered single-service benchmarks, \restberus produces statistically equivalent evaluation results to \restgym across all five metrics, with confidence intervals entirely within the predefined equivalence margins.
This demonstrates that our extended benchmarking infrastructure preserves comparable tool assessments on single-service applications.
\end{resultbox}

\subsection{Threats to Validity}

\textit{Internal Validity} concerns the extent to which the observed effects are caused by experimental factors rather than uncontrolled variables.
To mitigate these threats, all tools ran under identical conditions with a request budget instead of a time limit, limiting the effect of cluster throughput.
Moreover, we isolated each run in a dedicated Kubernetes namespace and performed ten repetitions with controlled random seeds.
Incomplete or stalled runs were retried with identical parameters.

\textit{Construct Validity} concerns whether the adopted evaluation measures accurately capture the intended concepts.
We consider HTTP 5xx responses as potential failures and estimate failure uniqueness through deduplication.
However, determining unique failures is challenging in black-box REST API testing, and manually inspecting tens of thousands of failure pairs is infeasible.
To assess automated deduplication accuracy, we used reference labels from a multi-model LLM judge ensemble, mitigating potential labeling errors through model diversity, zero-temperature prompting, and validation against 346 manually annotated pairs ($\kappa = 0.758$).
For failure attribution, \restberus requires trace evidence to classify a failure as downstream; otherwise, it conservatively attributes it to the target service.
This avoids unsupported downstream attributions, although missing or incomplete spans may cause some downstream manifestations to be overlooked.

\textit{External Validity} concerns the generalizability of our findings.
We selected six testing tools representing distinct generation paradigms across two distributed benchmark systems.
Both benchmarks use Java and Spring, which limits generalizability to polyglot architectures or alternative communication protocols, but they differ in size and dependency structure.
Moreover, circuit-breaker fallbacks in Train Ticket were disabled so that downstream failures remain observable at the API boundary.
Compared with a production deployment that retains such resilience mechanisms, more downstream failures are visible to the tools, so our downstream-failure ratios characterize what tools can expose in hardened systems rather than replicate production conditions.

\textit{Conclusion Validity} concerns the degree to which our statistical analysis supports our conclusions.
To account for stochastic test generation, we repeated each execution ten times~\cite{ArcuriAndrea2011Apgf} and used paired non-parametric tests.
For RQ1, we compared deduplication strategies using McNemar's test and corroborated LLM-based results with human annotations.
For RQ3, we did not infer equivalence from non-significant differences alone, but additionally required the 90\% confidence intervals of paired differences to lie entirely within predefined practical equivalence margins.

\section{Lessons Learned and Actionable Insights}

\textbf{Benchmark complete service chains, not isolated APIs.}
Benchmarking testing tools exclusively on isolated services overlooks failure modes arising from interactions with real downstream dependencies, e.g., cascading timeouts or downstream crashes.
Across our distributed benchmarks, all six testing tools exposed downstream failure manifestations, accounting for 11.7\% of unique failures on average.
Benchmarking campaigns should complement single-service evaluations with deployed microservice systems and assess failures across service boundaries.

\textbf{Look beyond response bodies and API operations to distinguish failures.}
Similar HTTP error responses may correspond to different failures, while the same downstream failure may be triggered through different API operations.
Our \texttt{hybrid} deduplication strategy combines traces and response bodies when traces are available, and operation-based grouping otherwise, achieving 87.1\% deduplication accuracy compared with 74.1\% for the payload-only strategy adopted by \restgym and 77.4\% for the operation-only strategy adopted by MacroHive.
Benchmarking infrastructures must leverage inter-service execution evidence, such as distributed traces, to observe how requests propagate, distinguish target-service from downstream failures, and deduplicate failures accurately.

\textbf{Modernize benchmarking infrastructure without breaking existing tools.}
Supporting distributed applications requires more sophisticated deployment and observation mechanisms, but these changes should not alter the evaluation of existing tools on established benchmarks.
\restberus runs existing REST API testing tools without modification and achieves statistically equivalent results to \restgym.
Benchmarking platforms should preserve tool compatibility and established metrics, allowing researchers to build on existing tools and empirical results while extending their evaluation to more realistic distributed environments.

\section{Related Work}

\subsection{Benchmarking Automated REST API Testing}

REST API relevance has motivated several automated test generators, whose effectiveness is usually assessed through API operation coverage, code coverage, and observed failures \cite{GolmohammadiAmid2024TRAA}.
\restgym~\cite{CorradiniDavide2025RAFI} facilitates reproducible comparisons of black-box testing tools by providing containerized benchmark applications, standardized execution, and common evaluation metrics.
However, its benchmarks target individual services, and its failure deduplication relies only on response-body similarity.
Our approach extends controlled tool comparisons to deployed microservice dependencies and incorporates distributed traces to improve failure analysis, while preserving comparable results on single-service benchmarks.

Other work offers experimental resources for evaluating REST API testing.
Gregor et al.~\cite{GregorLena2025BCaI} provide manually developed component- and integration-level test suites and a fault dataset for Train Ticket.
Defects4REST \cite{mehta2026defects4rest} offers a benchmark of real-world REST API defects, enabling effectiveness studies against known defects.
WFC/WFD \cite{sahin2025wfc} provides reusable API subjects and common specifications for authentication, fault reporting, and experimental comparisons.
These contributions are complementary to \restberus, which focuses on reproducibly executing test generators against deployed microservice systems and measuring their effectiveness through consistent, trace-aware analysis.

\subsection{Microservice Testing and Runtime Evidence}

MacroHive \cite{GiamatteiLuca2024AFaR} combines automated test generation with service monitoring and causal analysis to expose failures across microservice dependencies, including internal failures masked at the API boundary.
It provides an asymmetrical comparison with existing tools, as it collects internal failure information for MacroHive while evaluating the other tools through API-boundary metrics.
Unlike MacroHive, \restberus applies the same trace-aware failure analysis procedure to every evaluated tool, enabling comparisons of target and downstream failures without changing test generation logic.
Our approach also differs in scope: MacroHive can observe internal failures independently of the entry-point response, whereas \restberus analyzes failures associated with observed HTTP 5xx responses.

Other approaches exploit runtime observations to improve microservice testing.
MicroFuzz \cite{DiPeng2024MAEF} combines distributed tracing with service mocking to support efficient fuzzing, while MISH \cite{CaoClinton2025ATGf} learns from microservice logs to guide EvoMaster's test generation.
Zhang et al.~\cite{ZhangMan2025FMAS} also demonstrate the application of EvoMaster to industrial microservice systems.
Instead of introducing another generation technique, \restberus evaluates existing black-box generators under common experimental conditions.

Distributed traces have also been used for other testing purposes.
Abdelfattah et al.~\cite{AbdelfattahAmrS2024TCiM} use traces to measure microservice endpoint coverage, while Rios et al.~\cite{ibm_cloud22} use them for fault localization.
Instead, \restberus uses traces to distinguish failures manifesting in target and downstream services and to improve failure deduplication, enabling more informative comparisons of REST API testing tools.

\section{Conclusions and Future Work}

We presented \restberus, a Kubernetes-based benchmarking platform for evaluating black-box REST API testing tools on microservice systems.
\restberus introduces a failure analysis strategy that leverages distributed traces to distinguish target-service failures from downstream failures and to deduplicate them.
Across six tools, our novel deduplication strategy achieved 87.1\% deduplication accuracy, compared with 74.1\% and 77.4\% for payload-only and operation-only strategies proposed in existing work.
Our assessment considered failures wherever they manifest: all testing tools exposed both target-side and downstream failures, with the latter accounting for 11.7\% of unique failures on average.
\schemathesis discovered the most unique failures overall.
Our study demonstrates that benchmarking can move beyond API-boundary observations without sacrificing compatibility with existing tools or comparability with prior results.

Future work will extend \restberus to support real-world polyglot systems, including Go and Python.
We also plan to expose runtime traces as feedback for microservice-aware test generation.

\balance

\bibliographystyle{IEEEtran}
\bibliography{references}

\end{document}